\def\year{2020}\relax
\documentclass[letterpaper]{article} %
\usepackage{aaai20}  %
\usepackage{times}  %
\usepackage{helvet} %
\usepackage{courier}  %
\usepackage[hyphens]{url}  %
\usepackage{graphicx} %
\usepackage{graphicx}  %
\usepackage{tabularx}
\usepackage{balance}
\usepackage{booktabs}
\usepackage{amsmath}
\usepackage{wrapfig}

\usepackage{float}
\usepackage{tikz} 
\usepackage{xcolor}

\title{Leveraging Clickstream Trajectories to Reveal Low-Quality Workers in Crowdsourced Forecasting Platforms}

 \author{ Akira Matsui,\textsuperscript{1,2}
Emilio Ferrara,\textsuperscript{1,2,3}
Fred Morstatter,\textsuperscript{1,2}
Andr\'es Abeliuk,\textsuperscript{1,2}
Aram Galstyan\textsuperscript{1,2}\\
\textsuperscript{1}{University of Southern California, Information Sciences Institute}\\
\textsuperscript{2}{University of Southern California, Department of Computer Science}\\
\textsuperscript{3}{University of Southern California, Department of Communication}\\
}

\begin{document}

\maketitle

\begin{abstract}
Crowdwork often entails tackling cognitively-demanding and time-consuming tasks. Crowdsourcing can be used for complex annotation tasks, from medical imaging to geospatial data, and such data powers sensitive applications, such as health diagnostics or autonomous driving. 
However, the existence and prevalence of underperforming crowdworkers is well-recognized, and can pose a threat to the validity of crowdsourcing. 
In this study, we propose the use of a computational framework to identify clusters of underperforming workers using clickstream trajectories. 
We focus on crowdsourced geopolitical forecasting. The framework can reveal different types of underperformers, 
such as workers with forecasts whose accuracy is far from the consensus of the crowd, those who provide low-quality explanations for their forecasts, and those who simply copy-paste their forecasts from other users. Our study suggests that clickstream clustering and analysis are fundamental tools to diagnose the performance of crowdworkers in platforms leveraging the wisdom of crowds.
\end{abstract}

\section{Introduction}

Crowdsourcing has found applications in a variety of prediction domains, spanning politics, economics, technological and social issues~\cite{kittur2008crowdsourcing,kittur2011crowdforge}.
The wisdom of the crowd provides a powerful framework to tackle complex estimation problems, such as forecasting~\cite{howe2006rise,brabham2013crowdsourcing}.
Furthermore, crowdworkers are often used in the research pipeline for tasks that involve annotation and validation of data, and more recently to carry out complex decision-making tasks, such as geopolitical forecasting~\cite{doan2011crowdsourcing}.

Although, by definition, the crowd has the ability to absorb  anomalies in the behavior of certain workers, in some application domains it has been shown that underperforming individuals can negatively affect the quality, and even the ultimate validity, of an estimate or forecast~\cite{schenk2011towards,yuen2011survey}.

For example, in the context of geopolitical forecasting~\cite{ungar2012good}, early incorrect predictions may influence the consensus, which influences the behavior of later forecasters. This process propagates initial errors, and can yield a final estimate that is severely inaccurate, or even the opposite of the correct answer~\cite{moore2016confidence,friedman2018value}. In other cases, workers may act in a ``parasitic'' fashion and add no valuable knowledge or work toward solving a task, for example by simply mimicking the crowd's decisions.

For such reasons, it is of paramount importance to be able to diagnose the performance of crowdworkers in platforms that leverage the wisdom of the crowd. In particular, the ability to reveal low-quality workers who are engaging in undesirable behaviors could lead to timely solutions that will ultimately benefit the overall quality of crowdsourcing, e.g., informing workers' training, behavioral interventions, and incentives redesign. 

In this work, we investigate the problem of detecting low-quality workers that engage with three types of undesirable behaviors: \textit{(i)} producing forecasts that are unreasonably (and incorrectly) far from the consensus of the crowd at the time of prediction, or far from the final mean estimate; \textit{(ii)} simply copying the forecasts of other users or  adopting the consensus estimates of the crowd without providing any further information; and, \textit{(iii)} providing low-quality explanations to the reasoning that led them to produce their estimates. 

\subsection{Contributions of this work}
This work makes the following novel contributions:
\begin{itemize}
    \item Characterising the problem of identifying low-quality crowdworkers without having any ground truth for the assigned tasks, providing a model that can easily be applied to other crowdwork tasks.
    
    \item Conducting empirical evaluations with rich data from the \texttt{ANONYMOUS} platform we operate, that is clickstream trajectories, \textit{i.e.,} sequences of actions that took place on the platform generated by 547 Amazon Mechanical Turkers.
    
    \item Providing evidence that our proposed method can identify under-performing crowdworkers: the identified group of workers shows low-performance across the three different case studies we present.
\end{itemize}

\section{Study Design \& Data}
In this section, we discuss the design of our study, providing information on our geopolitical forecasting platform, details about the crowdsourced prediction tasks, and description of the data at hand.\footnote{This study received IRB approval and a research protocol from our institution. Details are removed for peer review purposes.}

\subsection{The forecasting platform}\label{sec:platform}

In this study, we leverage data from a platform we developed and operate, called \texttt{ANONYMOUS}.\footnote{Platform's name, acronym, and link are anonymized for peer review purposes.} 
\texttt{ANONYMOUS} is a forecasting platform for geopolitical events. Predictions are carried out by human users, who have also access to statistical models and data visualization tools to aid their work. A rich description of the \texttt{ANONYMOUS} platform is provided in our prior publications.\footnote{Citations to our prior work removed for peer review purposes.}

\subsubsection{Crowdworkers and HITs}
The whole population of \texttt{ANONYMOUS} forecasters studied in this work is constituted of crowdworkers. Specifically, we recruited Mechanical Turkers from Amazon Mechanical Turk (AMT).\footnote{Amazon Mechanical Turk: \url{https://www.mturk.com}}
Workers are paid to participate in our platform and to generate forecasts to the questions hosted on \texttt{ANONYMOUS}. In particular, each worker can complete one Human Intelligence Task (HIT) per week. Each HIT is constituted of answering to two new questions, and providing three forecast updates to previously-answered questions.

\subsubsection{Forecasting Problems}
In \texttt{ANONYMOUS}, questions (a.k.a., ``forecasting problems'') are posed in a multiple-choice format, consisting of the question text and a set of possible answer options. Importantly, all questions concern future events. The answer options consume the entire space of possible outcomes, separated into mutually-exclusive choices. The users can associate comments (a.k.a., \textit{rationales}) to their predictions. Figure \ref{fig:platform_image} (c) shows an illustrative example of a question with  five options: the users are required to set the probability of these outcomes. The user depicted in  Figure \ref{fig:platform_image} (c) set 20\% for each option and provided a (fictitious) comment to corroborate the rationale of their forecast. 

Users can also update their predictions, and associated rationales before a given question closes. Questions have end dates, after which they are resolved and scored. Forecasts are scored using an accuracy measure called Brier score, which rewards forecasts assigning the highest probability to the correct outcome out of $N$ options~\cite{brier1950verification}. Brier score is calculated as $\frac{1}{N} \sum^{N}_{i=1} (p_{i} - o_{i})^{2}$, where $N$ is the number of option, $p_i$ and $o_i$ the prediction and the outcome of option $i$, respectively.

After a question closes and resolves, we calculate the score for each user's prediction based on the actual outcome of that event. Hence, the Brier score is calculated for each user who participated to that question: the platform provides rewards and motivational affordances to incentivize accurate forecasting, such as badges, a higher position on the leaderboard, monetary incentives, etc.

\subsubsection{Charts \& Tools}\label{sub:consensus}
\begin{figure*}[t]
 \begin{center}
  \includegraphics[width=0.95\linewidth, height=0.4\linewidth]{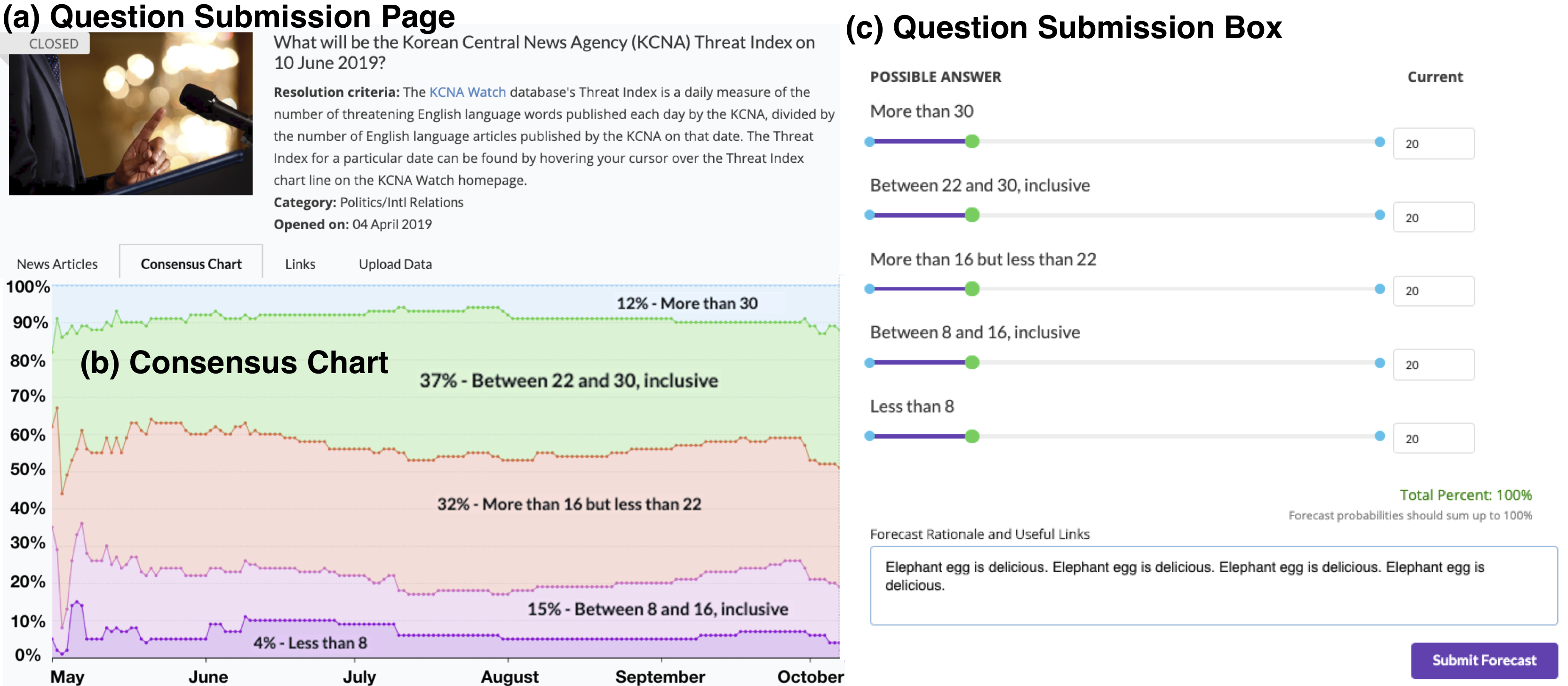}
  \caption{Example of a fictitious forecasting problem as it would appear on our \texttt{ANONYMOUS} platform. The forecasters select a question to answer in the platform. Each question has its \textbf{(a) Question Submission Page}  where forecasters see the explanations. On each submission page, \textbf{(b) Consensus Chart}  shows the average probability assigned to each outcome by all the forecasts produced on that question, up to any given point in time. While a question is open, forecasters can view the consensus trends in the answers of their peers, which may influence their decisions. For statistical purposes, Consensus Charts are not shown until 10 forecasts have been posted for that given question. On \textbf{(c) Question Submission Box} submitting forecast,  the forecaster is required to answer multiple-choice questions where the options are possible outcomes divided into mutually-exclusive bins. Forecasts are probability assignments for each outcome. A rationale in a free-form text can also be provided by the user to corroborate their forecast.}
  \label{fig:platform_image}
 \end{center}
\end{figure*}

Besides the forecasting task, users can interact at their own discretion with multiple tools in the platform in order to aid them in the forecasting process. These different tools, for example, allow users to: \textit{(i)} see other users' forecasts and rationales; \textit{(ii)} display the consensus chart showing the average forecast over time (see Figure~\ref{fig:platform_image} (b)); \textit{(iii)} recommend relevant news articles; \textit{(iv)} display charts showing relevant historical data; and, \textit{(v)} allow users to interact with different statistical models that produce forecasts based on the historical data.

\subsection{Data description and statistics}

Forecasting activity on the \texttt{ANONYMOUS} platform started on April 3, 2019, following a brief period of recruitment on AMT to identify a suitable pool of  participants. For this study, the last day of activity recorded in the data is November 29, 2019, accounting for exactly 6 months of activity records on \texttt{ANONYMOUS}. In such a period, 547 users produced over 670 thousand actions, recorded by the server-side back-end activity log of our system, which relies on a custom-made logging engine that captures hundreds of browsing and clickstream actions. For the purpose of this study, we focus on 22 main-category actions, and 45 subcategory actions (illustrated in Table~\ref{tab:clickstream_data}). The users produced over 56 thousand total forecasts, of which nearly 30 thousand were new forecasts (first-time forecasts on a given question) and the rest were updates to prior forecasts, across 410 forecasting problems.

\subsubsection{Action Logs}\label{sub:actions}

We built a custom, sophisticated activity logging system that tracks and records a large number of browsing and clickstream actions that each user performs on the Web client side. The system observes the users actions by monitoring the page elements that the user has active (i.e., rendered on the screen), and the buttons, tabs, and other user interface components that the user clicks. Despite not reporting the complete list of action logs, it is worth noting that over 95\% of the recorded action patterns are captured by just the top 10 actions, as portrayed by Table~\ref{tab:top_actions}.

\begin{table}[t]
    \centering
    \caption{Top 10 actions, which cumulatively account for about 95\% of the observed patterns on the \texttt{ANONYMOUS} platform.}
    \footnotesize
    \begin{tabular}{@{}llrr@{}}
    \toprule
    Category & Subcategory &   Count &  Total \%  \\
    \midrule
     View (HIT instructions)  &  &291,345 &  49.22\% \\
    Consensus chart &        View &   76,126 &  12.86\% \\
           Forecast &      Create &   73,227 &  12.37\% \\
      News articles &        View &   62,217 &  10.51\% \\
              Chart &        View &   49,427 &   8.35\% \\
              Links &        View &    5,127 &   0.87\% \\
   Resolution links &       Click &    4,608 &   0.78\% \\
             Filter &   Questions &    3,585 &   0.61\% \\
              Links &      Create &    3,425 &   0.58\% \\
      News articles &        Open &    3,066 &   0.52\% \\

    \bottomrule
    \end{tabular}
    \label{tab:top_actions}
\end{table}

\subsubsection{Clickstream data}
Clickstreams are sequences of actions that capture the workflow of a user. As we will demonstrate, clickstreams can be clustered to reveal common patterns of activity of workers on \texttt{ANONYMOUS}. Clickstream data allows us not only to account for the order of the actions but also for their duration. Also, question-level logs allow us to find if users copy-paste forecasts from the consensus chart by assessing whether they accessed it prior to forecasting and adopted the consensus estimates. We will discuss these strategies in the next section.

In Table~\ref{tab:clickstream_data} (d), we show an example of the structure of the clickstream data that we have access to from our \texttt{ANONYMOUS} platform. Each user's activity is recorded as a trajectory of actions, assigned with an ID (\textit{question\_id}, the associated user (\textit{user\_id}), a \textit{timestamp}. In this example, a forecaster (user\_id 1234) made the forecasts for the two questions (question\_id 1599 and 1581). After opening the first question, the user checked the consensus chart, then created a forecast and finally rated the difficulty of the question. Similarly, for the second question (question\_id 1581), the forecaster read the News articles and made the forecast. From these trails, we can observe that the forecaster spent more time on the latter question, after accessing the news articles.

\begin{table}[t]\small
    \caption{Example of structure of clickstream data for actions (a\_id), users (u\_id) and questions (q\_id).}
    \footnotesize
        \begin{tabular}{@{}l@{}@{}llll@{}r@{}@{}}
        \toprule
        a\_id & u\_id &   timestamp &         category & subcat. &  q\_id \\
        \midrule
        1   &   1234    &  10.4.2019 9:02:11 &             View &             &         1599 \\
        2   &   1234    &  10.4.2019 9:03:39 &  Consensus chart &        View &         1599 \\
        3   &   1234    &  10.4.2019 9:04:31 &         Forecast &      Create &         1599 \\
        4   &   1234    &  10.4.2019 9:04:32 &           Rating &             &         1599 \\
        5   &   1234    &  10.4.2019 9:05:11 &             View &             &         1581 \\
        6   &   1234    &  10.4.2019 9:05:22 &    News articles &        View &         1581 \\
        7   &   1234    &  10.4.2019 9:12:55 &         Forecast &      Create &         1581 \\
        \bottomrule
        \end{tabular}

    \label{tab:clickstream_data}
\end{table}

\section{Methods}

In this section, we describe our proposed methodology. First, we formalize the research question of identifying low-quality workers, and then we present the details of the proposed clickstream clustering framework. Finally,  we will discuss the methods to assess low-quality forecasts.

\subsection{Identification of low-quality workers}\label{sub:lowqual}
In this section, we postulate the following research question: 
\emph{\textbf{Can we identify low-quality workers from their behavioral trails?}}
We suggest its implications for crowdwork in geopolitical forecasting and broader applications, and propose a possible solution framework.

\subsubsection{Broader applications}
In other settings, it may be even harder to asses participants' performance. For example, if crowdworkers are used to answer surveys, which is commonly done in social science research~\cite{sheehan2018crowdsourcing}, it would be hard to distinguish between a truthful answer or otherwise not, solely based on the observed evidence, in part because of the lack of background information on the worker, and in part because these answers may carry no correlation with prior observed performance or behavior.

\subsubsection{Proposed solution}
A solution to this conundrum is to refrain from relying on observed past and current answers, but rather trying to codify the behavior of the workers from the digital traces of their activity on the crowdsourcing platform~\cite{lu2005mining}. In particular, in our work, we suggest using data that is not directly related to the participants' past forecasting performance, but rather to identify patterns of their behavior that reflect their expertise, commitment, and engagement to the forecasting task.

To that aim, we propose to cluster the participants based on their behavior on the platform to reveal low-quality workers. We assume that low-quality behaviors can be separated from valuable forecasting behaviors. Hence, we propose to use a clustering framework to find  users who exhibit consistent patterns of behavior to assess whether these associate with low-quality (or valuable) work. We view quality from several perspectives throughout the course of the paper, including forecasting ability and effectiveness of communication. From each perspective, we measure how well clickstreams can approximate the quality of the worker.

\subsubsection{Quality metrics}
After we cluster the users, we will investigate the prevalent behaviors in each cluster as conveyed by the clickstream activity.  If we observe that a cluster shows low-quality work according to a single criterion, we won't necessarily conclude that the workers in that cluster are low-quality. In fact, low-quality results according to a single metric may be misleading, as the workers in the incriminated cluster may produce valuable work with respect to other dimensions. To avoid this Type I error, we will study the quality of workers from different points of view (i.e., by employing three definitions of work value), and see if there is a cluster that displays consistent low quality. For validity assessment, we explore the forecasts from three perspectives: (i) the distance of forecasts from the total mean of the crowd, (ii) the degree that users copy others' estimates, and (iii) the readability of the rationales.

\subsection{Clickstream trajectory clustering}\label{sub:clickstream}

\begin{figure}[t]
  \centering
  \includegraphics[width=1.1\columnwidth]{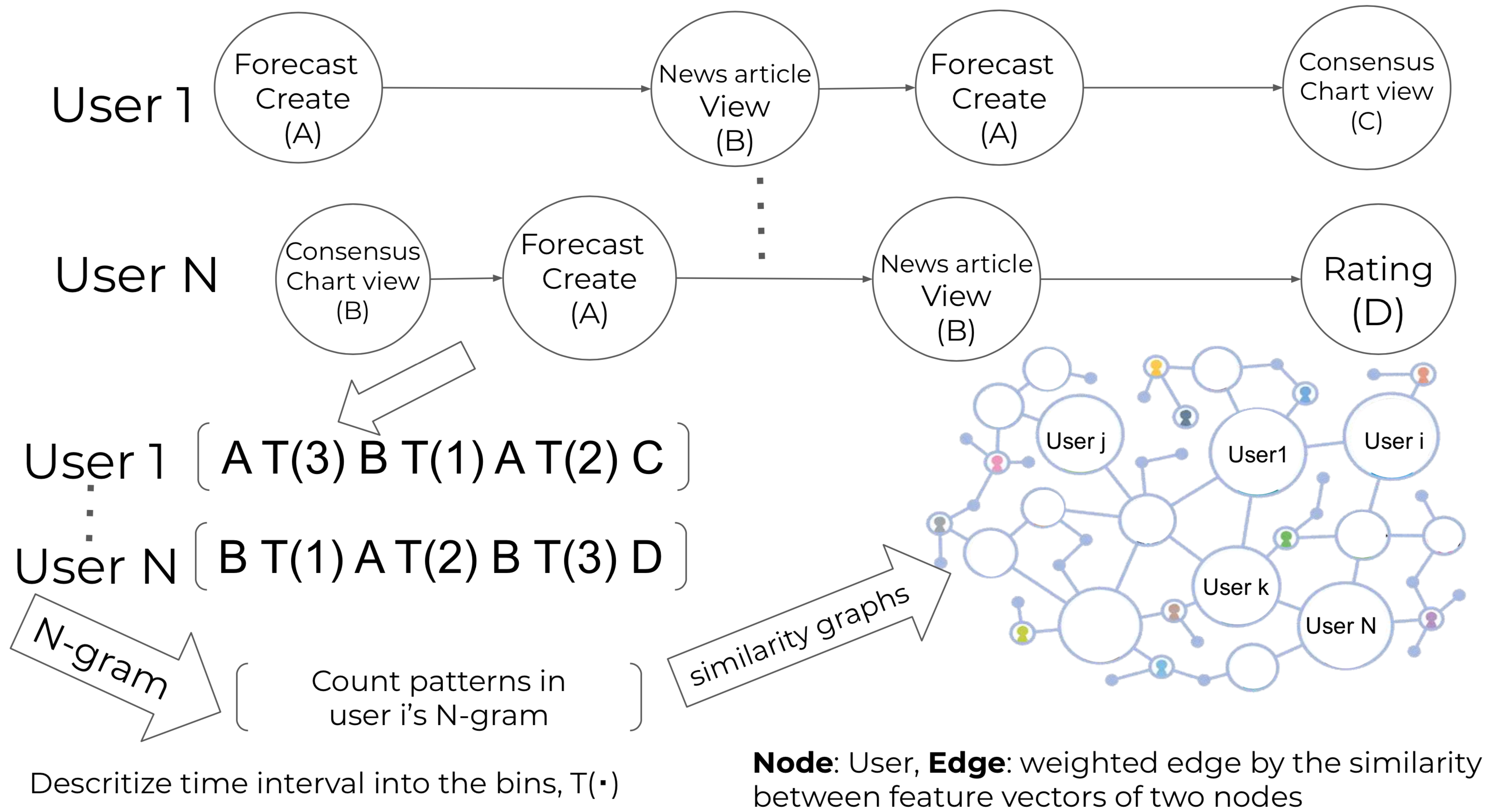}
  \caption{Illustration of our framework: from clickstream trajectories to similarity graph clustering.} 
  \label{fig:clickstream_clustering_illust}
\end{figure}

To cluster the participants based on their behavior in the platform, we focus on the clickstream, a sequence that describes how a user navigates and clicks on items on the Web platform~\cite{wang2016unsupervised}. %
To find the users who share behavioral patterns of clickstream trajectories, we utilize the unsupervised clickstream clustering approach inspired by Wang and collaborators~\cite{wang2016unsupervised}. This method finds clusters of similarly behaving users, where similarity is defined at the level of clickstream sequences. We provide an overview of the procedure of the clickstream clustering in Figure~\ref{fig:clickstream_clustering_illust}. 

In the clustering procedure, we first represent the clickstream trajectory as a sequence, $C_{i}$, of actions and discrete time intervals for each user $i$. Then, we count the occurrences of each possible n-gram within each sequence to form a feature vector $F_i$ of user $i$'s behavior.  Feature vector $F_i=\{f^i_j|\text{the number of element $f_j$ in $i$'s k-gram}\}$ represents the click patterns for user $i$.  %
With the feature vectors of users' click stream,  we construct the user similarity graph, $G=(V, E)$, where each node $v \in V$ correspond to user and edge is a weight calculated by similarity of the two node. In our similarity graph, a user similarity between user $i$ and $j$ is defined as an edge $v_{ij} \in V$ weighted by the polar distance (angular distance) of feature vectors $F_i$ and $F_j$,

\begin{align}
\begin{split}
PolarDist(F_i, F_j)&\\
=\frac{1}{\pi} \cos ^{-1} \frac{\sum_{k=1}^{n} f_{i k} \times f_{j k}}{\sqrt{\sum_{k=1}^{n}\left(f_{i k}\right)^{j}} \times \sqrt{\sum_{k=1}^{n}\left(f_{j k}\right)^{j}}}.
\end{split}
\end{align}

Finally, we cluster the nodes in the network using Divisive Hierarchical Clustering. By looking at the distribution of the elements in the feature vectors, we can study how a cluster $V_i \in V$ is different from other users not in cluster $V_i$. For example, for cluster $V_i$, we get the distribution of each element $f^i_j$ each user $i$ has in their k-gram count $F_i$. Then, we can compute the $\chi^{2}$ statistics to find the difference between cluster $V_i$ and the others. This $\chi^{2}$ can be interpreted as how different $V_i$ is from the other clusters.

\subsection{Assessment of low-quality forecasts}\label{sub:assess}
We examine the forecasts from four points of view. (i) Firstly, we assess the variance of the forecasts in each cluster. We calculate the root mean square error (RMSE) from the total mean.  (ii) Then we analyze the similarity of the forecasts to the consensus. Then, we detect copy-paste behavior. (ii) Lastly, we study the comments that the forecaster wrote as rationales accompanying their forecast. 

\subsubsection{Readability Score}
On our platform, users can make predictions for questions with comments as rationales, explaining the reason why the user did reach that prediction. Therefore, these comments might contain crucial information about their prediction behaviors. Comments suggest how much the users invested in terms of  time and effort to reach a decision. We computationally evaluate how much these comments are concise and readable. 

We calculate the readability scores to assess the quality of the comments, a notion first introduced by Klare~\cite{klare1963measurement}. However, the validity of the original readability score formulation by Klare is  controversial \cite{redish2000readability}. Therefore, in this paper, we use two other readability scores to look into the comments from different angles. We use \textit{Coleman-Liau Index} (CLI) \cite{coleman1975computer} and \textit{Automated Readability Index} (ARI) \cite{senter1967automated}. These two different scoring mechanisms have different goals \cite{liu2015makes}.  While ARI measures the reading ease, CLI measures the complexity of the text. In both scorings, the readability of the text decreases as its score increases. Equations \ref{eq:CLI} and \ref{eq:ARI} show the formulas for the two scores:

\begin{align}
CLI &=& 5.89\frac{ch}{wo} - 0.3\frac{se}{wo} - 15.8, \label{eq:CLI}
\\
ARI &=& 4.71 \frac{ch}{wo} + 0.5\frac{wo}{se} - 21.43,
\label{eq:ARI}
\end{align}
where $ch$ is total number of characters in a document; $wo$ is total number of words in a document; $se$ is total number of sentences in a document.
\subsubsection{Quantifying the distance of forecasts from the mean}

To find inconsistent forecast behavior, we calculate how far each forecast is from the mean. When a user generates a thoughtless forecast, that forecast will change the total mean. The magnitude of that change can be drastic early in the forecast life-span when the sample size of forecasts is smaller---hence, it may influence future forecasters and affect the validity of the whole prediction.
Accordingly, the variance of low-quality forecasts would also be higher than that of well-researched forecasts, under the reasonable assumption that workers who use similar information to answer a question would reach similar conclusions.
We calculate the mean square error from the total mean at each question. Then, we compare the mean and variance among the clusters. The root mean square error (RMSE) of the forecast $f$ made for a given question is calculated as

\begin{equation}\label{eq:RMSE_fq}
\operatorname{RMSE}(f) =\sqrt{\frac{1}{N} \sum_{j=1}^{N}\sum_{o=1}^{O}\left(f_j^o-m^o\right)^{2}},
\end{equation}
where $N$ is the number of forecasters, $O$ is the number of options for the given question, and $m^o$ is the mean value of option $o$ for the given question.
We first obtain $f$ for all questions for each cluster. Then, we calculate $\operatorname{RMSE}(f)$ to have the distribution of RMSE for each cluster.

\subsubsection{Quantifying the distance of forecasts from the consensus}

Next, we quantify how forecasts are similar to the consensus of the crowd. On our platform, we show a Consensus Chart associated to each question  (see Figure \ref{fig:platform_image} (b)-1). The consensus charts are updated daily and they show the distribution of the forecasts made by other workers chronologically, up until to the point in time when they are viewed.

It is worth noting that, while the total mean used in Equation \ref{eq:RMSE_fq} is the mean of all answers to a given question after it closes and resolves, 
the Consensus Chart shows the value of the mean forecast at the time when a forecaster viewed the chart (i.e., during the period when the question is open). 
The forecasters can see the trends of the crowd's consensus when they are forecasting. The conditions regarding the consensus chart are the same for all forecasters. After a question reaches 10 forecasts, the same consensus chart is available to all users. This consensus chart can provide the forecasters with a reference point and reduce the effort to make a prediction. On the other hand, the consensus charts may also  induce the forecasters to post a forecaster in line with the consensus chart, and sometimes even to copy and paste the consensus estimates, without adding any further valuable research work toward the question's answer. Incorporating the consensus into the forecast is legitimate. However, if some forecasters systematically rely only on the consensus, their forecasts will not add any information to the  consensus of the crowd, suggesting a parasitic and undesirable behavior in our platform. Hence, comparing the similarity of the forecasts to the consensus chart is a suitable strategy to find  low-quality work.

Similarly to the previous metric, we use again the RMSE to capture how each forecast is similar to the consensus chart for a given question, as follows:

\begin{equation}
\operatorname{RMSE_C}(f_{q}) =\sqrt{\frac{1}{O} \sum_{o=1}^{O}\left(f_{jo}-c_{o}\right)^{2}}\label{equ:rmse_c},
\end{equation}
where $N$ is the number of the forecaster, $O_q$ is the number of the options of question $q$, and $c_{o}$ is the consensus chart value for option $o$ of the question $q$ when user made their forecasts. We calculate the mean of $\operatorname{RMSE_C}(f_{i})$ across the all question $q$ that user $i$ answered,

\begin{equation}
AVG(\operatorname{RMSE_C}(f_{i})) =\frac{1}{|Q_i|} \sum_{q \in Q_i} \operatorname{RMSE_C}(f_{iq}),
\end{equation}
where $Q_i$ is the set of questions that user $i$ answered. Finally, we compare the distribution of $AVG(\operatorname{RMSE_C}(f_{i}))$ to see the differences among the clusters.

\subsubsection{Quantifying copy-paste behavior}
As an extreme case of referencing behavior to others' forecasts, we here examine copy-paste behavior. The availability of the consensus charts might tempt some users to copy the values of the consensus and paste them as their own forecast.  

When a forecaster posts an exact  copy of the consensus estimate, the error defined in Equation \ref{equ:rmse_c} is zero. Therefore, we can compute the ratio of copy-paste behavior of user $i$ as follow:

\begin{equation}\label{equ:copy_paste}
 Prob_i = \frac{1}{|Q_i|} \sum_{q \in Q_i} \mathbf{1} (\operatorname{RMSE_C}(f_{iq}) = 0),
\end{equation}

where $Q_i$ is the set of questions that user $i$ answered. Also, we will relax the criteria for copy-paste behavior. Even if a forecaster intents to \textit{de facto} copy the consensus estimate, the values they may actually post might be slightly different from the chart. This is because in the forecast submission box (see Figure~\ref{fig:platform_image} (c))  the forecasters  can also set their estimates by sliding the bars rather than by typing in numerical values Figure~\ref{fig:platform_image} (c)).  For example, when the consensus chart has (20\%, 20\%,  20\%,  20\%, 20\%) as its values, the forecaster might move the sliders to (20\%, 20\%,  20\%,  21\%, 19\%). It can  happen that the ``copy-paste'' forecast slightly deviates from the consensus value.
Therefore, we relaxed the threshold of the error to classify copy-paste behavior from zero to some threshold. Based on this insight, we re-define the function $ Prob_k(x)$ for cluster $k$ defined as follow

\begin{equation}
 Prob_i(x) = \frac{1}{|Q_i|} \sum_{q \in Q_i}  \mathbf{1}(\operatorname{RMSE_C}(f_{iq}) \leq x),
\end{equation}

where $x$ is the error threshold for copy-paste behavior. In the next section, we will analyze the distribution of $Prob(x)$ values by users. %

\section{Results}

Here, we present the clustering results, and then discuss three scenarios, namely the dispersion of forecasts, the detection of copy-paste behavior, and the detection of low-quality rationales.
After we present our results, we test an alternative explanation for the clustering results. 
\subsection{Clustering results}\label{sub:clustering}

With clickstream clustering, we cluster forecasters who share the same patterns. We use 5-grams to capture clickstream trajectories. Table~\ref{table:cluster_results} demonstrates that the clickstream clustering generates 19 clusters. The clickstream clustering identifies three large user clusters in which the users share common dynamic patterns. These three large clusters (\textsl{\textsl{Cluster 1}, 2, and 3}) account for more than half the population, whereas the rest of 16 clusters are much smaller. In other words, \textsl{Cluster 1, 2}, and \textsl{3} consists of the users collectively behave similarly. The rest of the clusters have diverse behavioral patterns. To have a balanced comparison group, we combine these small clusters and name this compounded clusters as \textsl{Cluster 4}. \textsl{Cluster 4} can be interpreted as users who do not share common action patterns. Our hypothesis is that either \textsl{\textsl{Cluster 1},2, or 3} potentially represents low-quality workers. In the following, we validate that hypothesis based on three scenarios concerned with low-quality crowdworkers.

\begin{table}[t]\footnotesize
    \caption{Clickstream clustering results}
    \label{table:cluster_results}
\begin{tabular}{@{}c@{}@{}l@{}cc@{}}
\toprule
Cluster\ \  & \ \ \   Clickstream trajectory &     $\chi^2$ & Size\\
\midrule
1       &    news-articles T[3] forecast\_create T[2] rating &   147.64 &     104 \\
2       &             rating T[2] consensus-chart T[2] view &  1070.52 &     133 \\
3       &             rating T[1] consensus-chart T[2] view &   193.99 &     148 \\
4       &                 view T[2] view T[1] news-articles &    74.00 &       5 \\
5       &                 view T[2] view T[1] news-articles &    42.00 &       3 \\
6       &                  forecast\_create T[2] rating T[2] view &  1170.95 &      18 \\
7       &                        view T[1] view\_recent T[2] view &    19.38 &       2 \\
8       &                 view T[2] view T[1] news-articles &    13.28 &       1 \\
9       &  forecast\_create T[2] rating T[2] consensus-chart &     8.77 &       1 \\
10      &                 view T[2] view T[1] news-articles &    49.14 &       4 \\
11      &                        view T[1] view\_recent T[2] view &   216.39 &      18 \\
12      &           chart T[2] view T[1] news-articles &  1864.60 &      33 \\
13      &    news-articles T[2] forecast\_create T[2] rating &    75.85 &       5 \\
14      &                               view T[2] view T[2] view &   551.71 &      26 \\
15      &    forecast\_create T[2] consensus-chart T[2] view &  1360.10 &      21 \\
16      &           view T[1] news-articles T[3] links &   672.63 &       8 \\
17      &              view T[1] chart T[2] forecast\_create &   300.67 &       6 \\
18      &                 view T[3] view T[2] news-articles &   458.73 &      16 \\
19       &                       links\_create T[2] view T[2] view &   656.40 &       6 \\
\bottomrule
\multicolumn{4}{l}{%
  \begin{minipage}{\columnwidth}%
    Each cluster is presented with the  highest $\chi^2$ score clickstream. $T$ represents an interval: $T[1]:< 1$ second, $T[2]: < 1$ minute, $T[3]: < 1$ hour, $T[4]: < 1$ day. The $\chi ^ 2$ score is calculated as the $\chi ^ 2$ between the cluster on that row and the others for the clickstream on the same row. The top three clusters are named as \textsl{Cluster 1}, \textsl{Cluster 2} and \textsl{Cluster 3} respectively. The rest of the others are grouped as \textsl{Cluster 4} (super cluster).
  \end{minipage}%
}\\
\end{tabular}
\end{table}
\subsection{Scenario 1: Dispersion of forecasts} \label{sub:dispersion}

\begin{figure}[t]
  \includegraphics[width=\columnwidth]{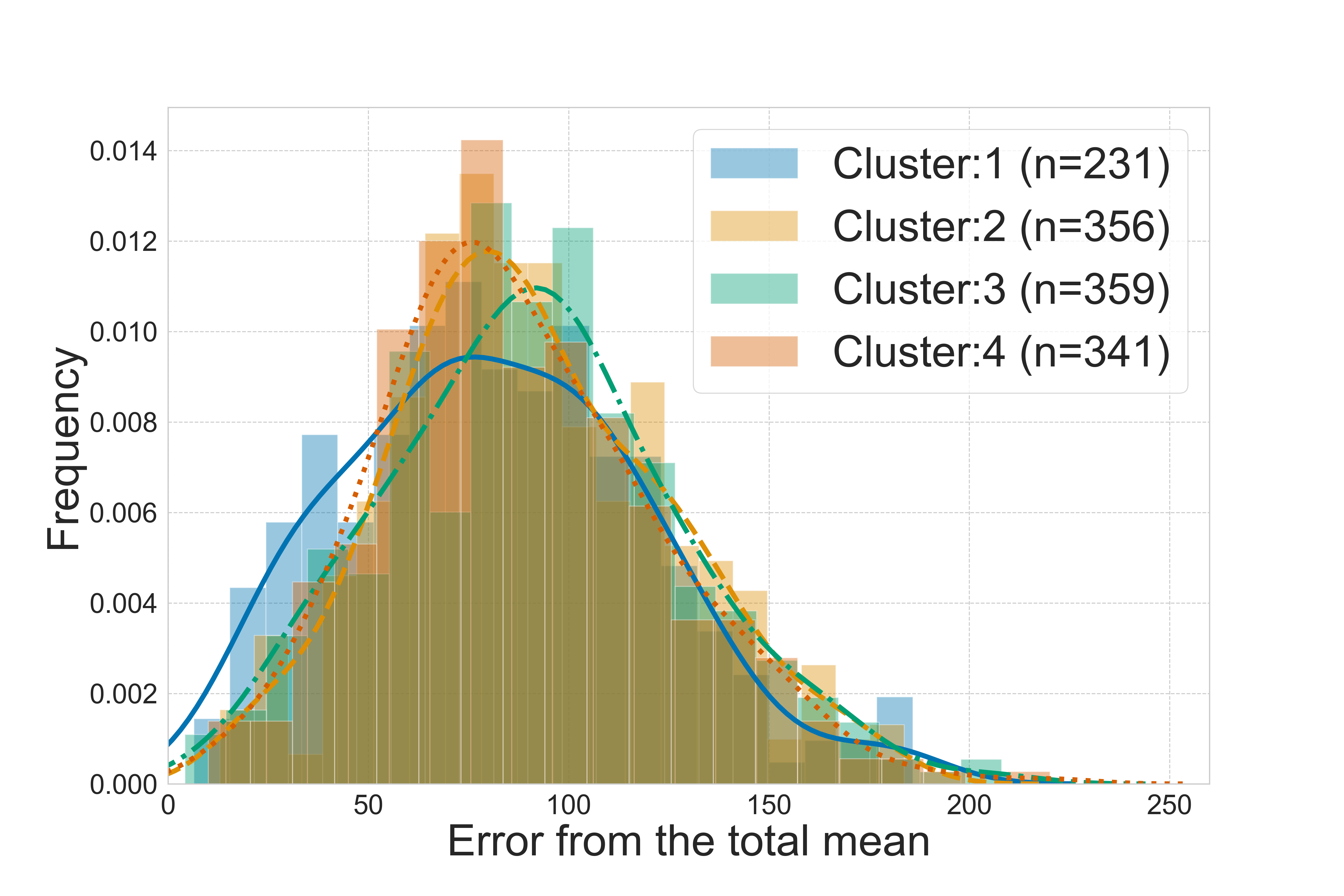}
  \caption{Distribution of the Root Mean Square Error (RMSE) between each user's answer and the mean of the answers of other forecasters on a given question, calculated for all the questions and partitioned by cluster. Solid lines: Kernel Density Estimate (KDE) for the observed data. } 
  \label{fig:error_from_mean}
\end{figure}

Low-quality forecasters might try to minimize the time and effort spent on each question. Their  forecasts would contain a larger random error than the others on average, hence generating a dispersed distribution (a different mean and larger variance than the distribution of the other forecasts). To study this scenario, we compare the means of forecasts across each cluster. We perform this analysis by comparing their difference from the total mean of the question. We calculate the $\operatorname{RMSE}(f_{q})$ (cf., Equation~ \ref{eq:RMSE_fq}) for each \textsl{Cluster } and plot the distribution in Figure \ref{fig:error_from_mean}. Figure \ref{fig:error_from_mean} shows that \textsl{Cluster 1} has smaller mean and higher variance than any other cluster. We present the mean values in Table~\ref{table:error_mean_value}. To compare these mean values, we conduct Welch's t-test and Levene's test to see if the clusters have different means and variances. \textsl{Cluster 1} has the lowest mean value and this difference is statistically significant (vs \textsl{Cluster 2} and \textsl{3}), $p < 0.01$; vs \textsl{Cluster 4}, $p \approx 0.057$). Notably, the null hypotheses of Welch's t-test for the other comparison are not rejected ($p > 0.17$). For variance, Levene's test for \textsl{Cluster 1} vs \textsl{Cluster 2} and \textsl{Cluster 1} and \textsl{Cluster 4}  are rejected at $p < 0.1$ but the other Levene's tests are not. These dispersed and inconsistent forecasts from \textsl{Cluster 1} tempts us to judge \textsl{Cluster 1} is low-quality forecaster cluster. To corroborate this hypothesis, we will study the clusters from two other angles in the following scenarios. 

\subsection{Scenario 2: Copying behavior} \label{sub:copypaste}

Low-quality forecasters can copy-paste the consensus chart which shows the mean estimates of the other forecasters, i.e., the crowd's consensus.  Copying-and-pasting the consensus chart is much easier than taking the time and energy to forecast by oneself.
In addition, this free-riding behavior will not reflect in a low performance as measured by the Brier score, as forecast scores will be close to the average forecast, hence it is particularly hard to detect. Notice how this type of behavior does not bring new information to the system, but, is not necessarily detrimental either. However, it is a waste of resources as crowdworkers are paid to work on these tasks.

To see if the forecasts leverage copy-paste behavior, we study the distance of the forecasts from the consensus chart. Note that the consensus chart is different from the total mean we used in Scenario 1. The consensus chart values are the weighted mean (with recent forecasts having more weight) of the others at the time the forecaster made the prediction, while we use the final mean in Scenario 1. Also, to be sure that the forecasts are made by copy-paste, we focus on the forecasts where the forecasters referred to the consensus chart before they made the forecasts.

\begin{figure}[t]
  \includegraphics[width=\columnwidth]{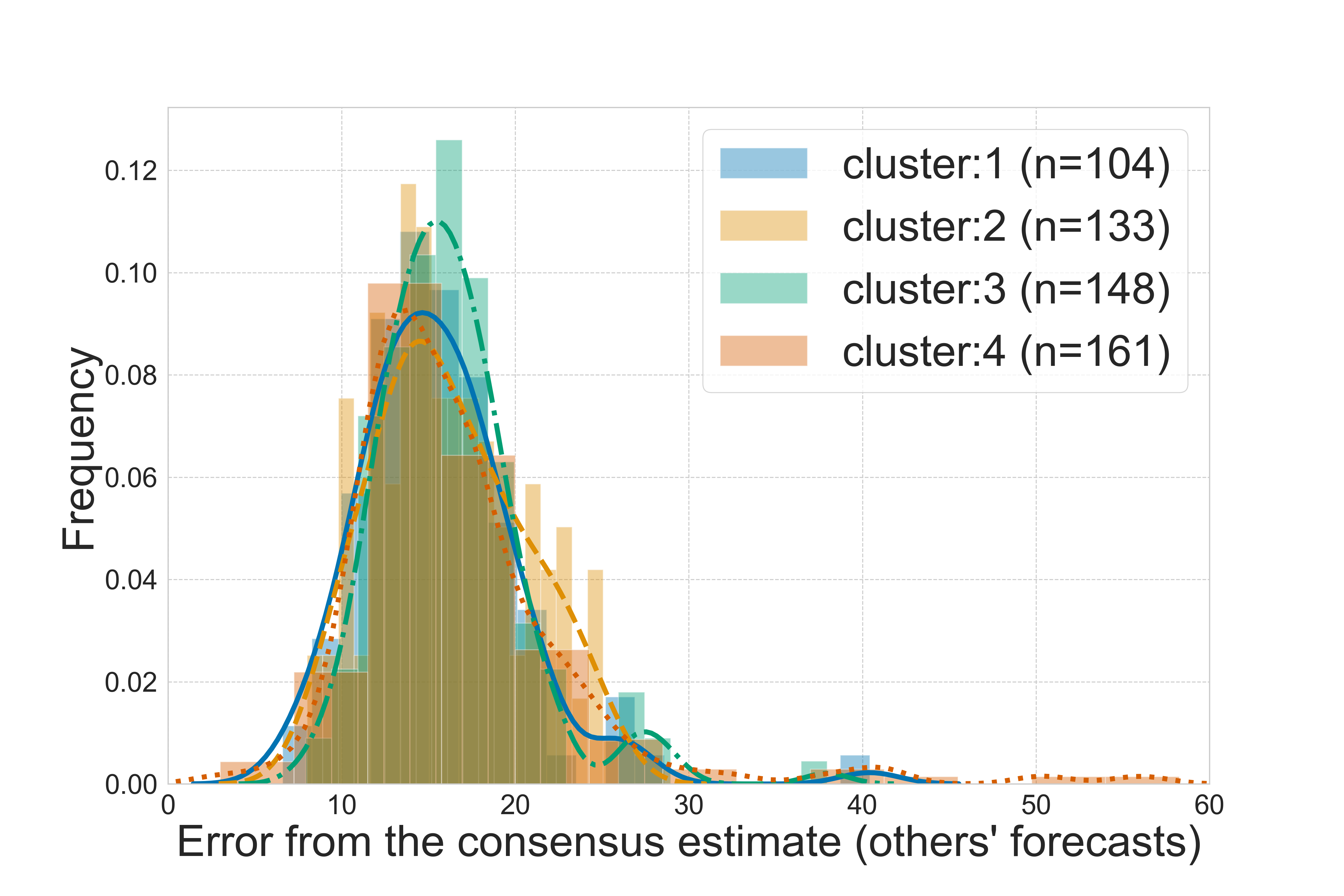}
  \caption{Distribution of the Root Mean Square Error (RMSE) between each user's answer and the consensus of the other forecasters on a given question, calculated for all the questions and partitioned by cluster. Solid lines: Kernel Density Estimate (KDE) for the observed data. } 
  \label{fig:error_from_consensus}
\end{figure}

The distribution of $\operatorname{RMSE_C}(f_{iq})$ in Equation \ref{equ:rmse_c} enables us to study how much the forecasters refer to the consensus chart at that time. The distributions of $\operatorname{RMSE_C}(f_{q})$ for each cluster are plotted in Figure \ref{fig:error_from_consensus} and the mean values of each cluster are presented in Table \ref{table:error_cc_stat}. The distribution differences among the clusters are subtle but \textsl{Cluster 1} has the smallest mean value. These differences are not statistically significant($p > 0.18$) except for the comparison between \textsl{Cluster 1} and \textsl{Cluster 4}($p \approx 0.068$). Even thought the differences are not definite, the crowdworkers in \textsl{Cluster 1} are the likeliest workers to copy-past other forecasts.

\begin{table}[t]
    \caption{Mean errors between answers and the overall mean}
    
\begin{tabular}{@{}l@{}@{}ccc@{}@{}}
\toprule
{}  &    RMSE: Mean &  Standard deviation &  \#Questions\\
\midrule
\textsl{Cluster 1} &  81.979 &              37.453 &           231\\
\textsl{Cluster 2} & 90.728 &              34.376 &          356\\
\textsl{Cluster 3} &  91.184 &              36.645 &           359\\
\textsl{Cluster 4} &  87.641 &              34.650 &           341\\
\bottomrule
\end{tabular}
    
    \begin{minipage}{\columnwidth}
      \small
       The RMSE is calculated as the error between each user's answer and the mean of each cluster. We compute the mean of RMSE for each cluster.
    \end{minipage}
 \label{table:error_mean_value}
\end{table}

\begin{table}[t] 
    \caption{Difference between forecasts and  consensus charts}
    
    \begin{tabular}{@{}l@{}@{}ccc@{}@{}}
    \toprule
    {} &  RMSE:  Mean &  Standard deviation  & \#Forecasters \\
    \midrule
    \textsl{Cluster 1} &  15.60 &               4.69 &          104 \\
    \textsl{Cluster 2} &  16.15 &               4.24 &          133 \\
    \textsl{Cluster 3} &  16.25 &               4.18 &          148 \\
    \textsl{Cluster 4} &  17.20 &               8.87 &          161 \\
    \bottomrule
    \end{tabular}
    \begin{minipage}{\columnwidth}\small
      The RMSE is calculated as the error between each  actual user's answer and the consensus of the other forecasters on a given question.
    \end{minipage}
    \label{table:error_cc_stat}
\end{table}

\begin{figure}[t]
  \includegraphics[width=\columnwidth]{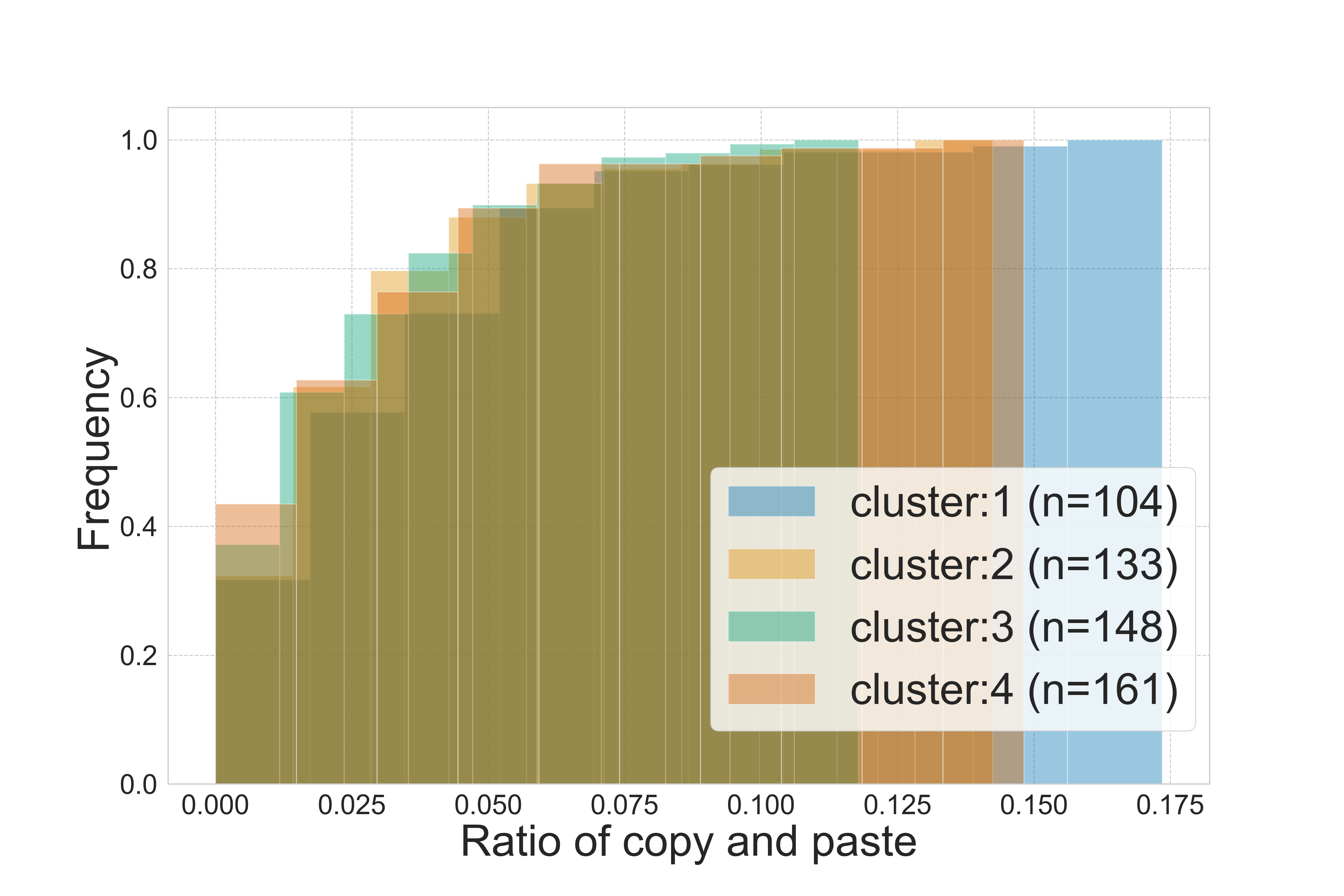}
 \caption{Cumulative Distribution Function (CDF) of the ratio of forecasts copy-and-paste  partitioned by clusters.} 
    \label{fig:copy_paste_dist}
\end{figure}

Now we focus on the probability of copy-paste behavior for each corwdworkers. This is because, even though we find that the forecasts in \textsl{Cluster 1} are more similar to the consensus chart than the others, we cannot judge that the forecasters in \textsl{Cluster 1} provide low-quality labor supply from that evidence alone. To verify that \textsl{Cluster 1} consists of low-quality forecasters, we study the ratio of copy-paste behavior, $Prob_i$, in each cluster. Figure~\ref{fig:copy_paste_dist} shows the cumulative distribution of $Prob_i$ (see Equation \ref{equ:copy_paste}) for each cluster. Figure~\ref{fig:copy_paste_dist} clearly shows that the forecasters in \textsl{Cluster 1} are copying from the consensus. In Table \ref{table:copycat_stat}, we show the user average of the ratio of the copy-paste forecasts to the total. We find that \textsl{Cluster 1} has higher copy-paste forecast probability than \textsl{Cluster 2} ($p < 0.05$), \textsl{Cluster 3}, and \textsl{4} ($p < 0.01$) whereas the other comparisons do not have a significant difference ($p > 0.40$). These results shows that the crowdworkers in \textsl{Cluster 1} are high likely to involved in copy-pasting from others' forecast.

\begin{table}[t]
    \caption{Average Ratio of copy and paste behavior}
    \begin{tabular}{@{}l@{}ccc@{}}
    \toprule
        {} &   Mean &  Standard deviation  &  \#Forecasters \\
        \midrule
        \textsl{\textsl{Cluster 1} }&  0.036 &               0.031 &          104  \\
        \textsl{\textsl{Cluster 2} } &  0.028 &               0.027 &          133 \\
        \textsl{\textsl{Cluster 3} } &  0.025 &               0.025 &          148 \\
        \textsl{\textsl{Cluster 4} } &  0.025 &               0.028 &          161 \\
        \bottomrule
\end{tabular}
    \begin{minipage}{\columnwidth}\small
      We count the number of forecasts having the exact same estimates as the consensus chart and compute their ratio to the total forecasts by each user. Then we compute the average ratio for each cluster.
    \end{minipage}
    \label{table:copycat_stat}
\end{table}

\subsubsection{Quasi copy-paste detection}
\begin{figure}[t]
\centering
  \includegraphics[width=0.75\linewidth]{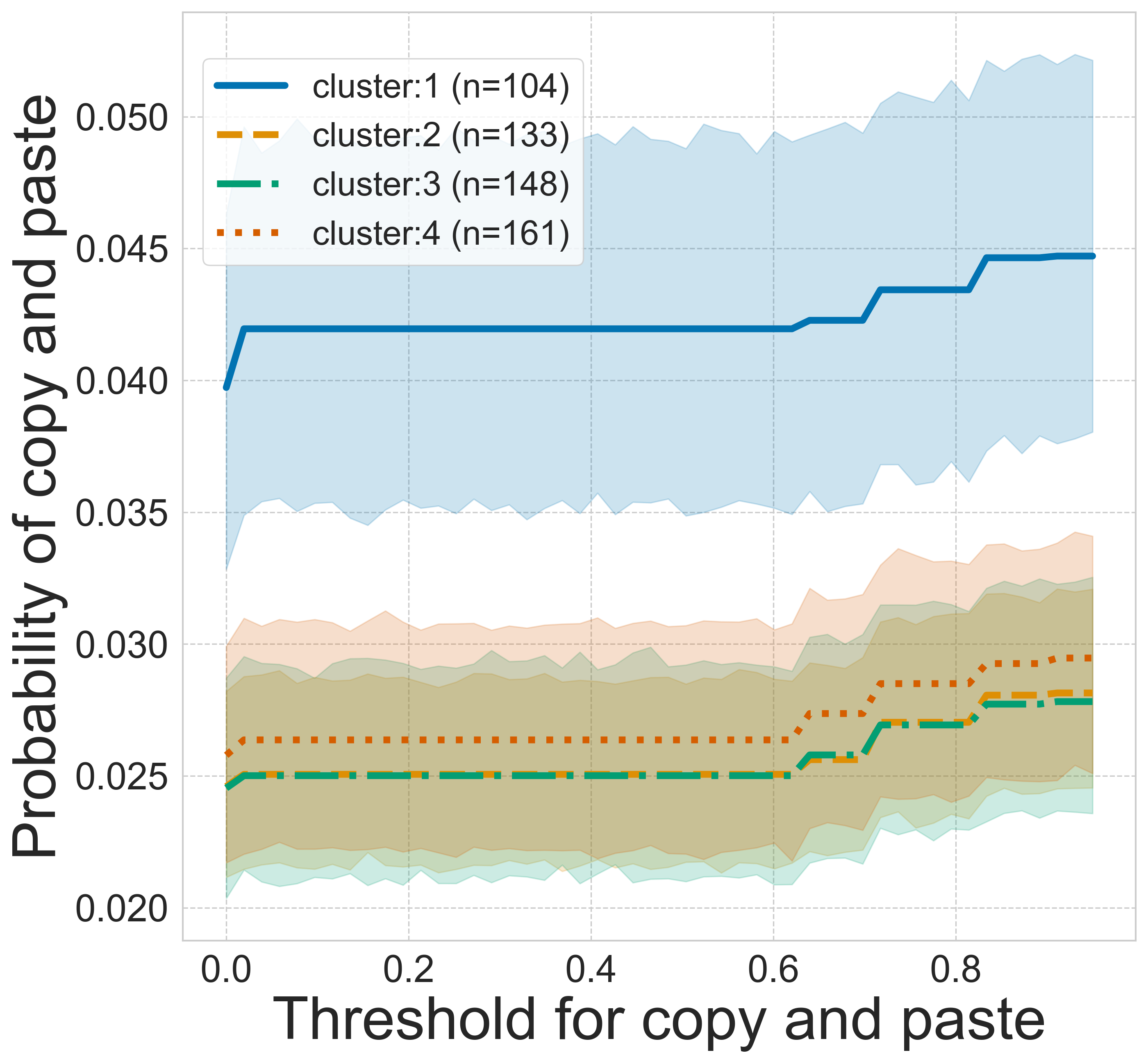}
\caption{The transition of the copy and paste per user across the range of the threshold for copy-paste behavior
(with 95\% confident intervals). We relax the threshold that is used for how much distance from the consensus chart is labeled as copy and paste. Across the entire figure, the copy-paste ratio of \textsl{Cluster 1} is higher than the other clusters.}
    \label{fig:thr_copy_paste}
\end{figure}

As the last evidence of copy-paste behavior, we relax the threshold of the copy-paste behavior to see the robustness of our results. We use the strictest criteria of copy-paste behavior in Figure \ref{fig:copy_paste_dist} and take the forecast that is completely copied and pasted from the consensus chart (i.e., the error from the consensus chart is zero). As we discussed in the previous section, copycats might not completely copy and paste the consensus chart.  We plot how the ratio of copy-paste forecast per user changes as we relax the threshold. Figure~\ref{fig:thr_copy_paste} shows our results are robust. Figure~\ref{fig:thr_copy_paste} clearly shows that \textsl{Cluster 1}'s ratio of copycats is higher across the whole range of thresholds than the other clusters.

\subsection{Scenario 3: Low-quality rationales}
\label{sub:rationales}
In our platform, the forecasters post a rationale to justify their forecasts. It is a very hard task to write a justification that is understandable to others. Therefore, low-quality forecasters may underperform on this task because writing good rationales does not directly reflect into the Brier scores. To study this point, we compare the quality of rationales. We assess the rationales for each cluster in three perspectives: length, the rate of misspelling, and readability. Before the analysis, we preprocess the raw rationale text to remove non-word tokens (e.g., URLs, and mentions of other users). All comments are written in English. 

The text length per rationale and average number of misspellings per rationale are presented in Table~\ref{tab:rational_length_misspel}. 
 \textsl{Cluster 1} writes the shortest rationale on average but the comparison with \textsl{Cluster 2} only show statistically significance differences ($p \approx 0.088$). As for misspelling per words in the rationale, all of the clusters has similar mean value ($p > 0.240$). 

\begin{table}[t]\small
    \caption{Mean values of text length and misspelling ratio}
       \begin{center}
    \begin{tabular}{@{}l@{ }ccc@{}@{}}
    \toprule
      {}&   Mean &  Standard deviation  &  \#Forecasters \\
    \textbf{Text length}  & & & \\
    \midrule
    \textsl{Cluster 1} &   69.97 &              24.46 &          104 \\
    \textsl{Cluster 2}  &  76.41 &              33.04 &          133 \\
    \textsl{Cluster 3} &  71.62 &              29.48 &          148 \\
    \textsl{Cluster 4}  &  71.27 &              36.396 &          162 \\
   \textbf{Misspell per word} &  & & \\
   \midrule
    \textsl{Cluster 1}   &   0.113 &               0.027 &          104 \\
    \textsl{Cluster 2}  &   0.112 &               0.026 &          133 \\
    \textsl{Cluster 3} &   0.112 &               0.027 &          148 \\
    \textsl{Cluster 4} &   0.117 &               0.051 &          162 \\
      \bottomrule
    \end{tabular}
   \end{center}
    \begin{minipage}{\columnwidth}\small
     We calculate the mean value of text length and misspelling. Misspelling is calculated as the number of misspelling divided by the total number of words in each user's rationale (comment). Both metrics are calculated for each user as an average. Text length is calculated as the number of words per rationale and the misspelling is the number of misspelling per word.
    \end{minipage}
    \label{tab:rational_length_misspel}
\end{table}

Table~\ref{tab:readability} describes ARI and CLI readability score. We do not see clear differences in ARI score($p > 0.13$). In CLI, \textsl{Cluster 4} has the smaller score than \textsl{Cluster 1}($p < 0.1$)  and \textsl{Cluster 2} ($p < 0.01$) but we could not reject the Welch's t-test for the comparison with \textsl{Cluster 3} ($p > 0.28$). We see other statistical significance differences between \textsl{Cluster 1} and \textsl{Cluster 3} ($p < 0.1$) , and  \textsl{Cluster 2} and \textsl{Cluster 3} ($p < 0.01$). Although the readability analysis on the rationales do not show clear results, the shortest and lower readability rationales by \textsl{Cluster 1} consists with the findings in the first two scenarios that \textsl{Cluster 1} is a low quality worker clusters.

\begin{table}[t] \small
\caption{Mean readability scores}
   \begin{center}
\begin{tabular}{@{}l@{}ccc@{}}
\toprule
{}&    Mean &  Standard deviation  &  \#Forecasters\\

    \textbf{ARI} & & &  \\
    \midrule
    \textsl{\textsl{Cluster 1}   }  &  33.39 &              12.02 &          104 \\
    \textsl{\textsl{Cluster 2}   } &  36.11 &              16.08 &          133 \\
    \textsl{\textsl{Cluster 3}   } &  33.55 &              14.37 &          148 \\
    \textsl{\textsl{Cluster 4}   } &   34.15 &              17.79 &          162 \\
    
    \textbf{CLI} & & &  \\
    \midrule
    \textsl{\textsl{Cluster 1}   } &  9.51 &               1.24 &          104 \\
    \textsl{\textsl{Cluster 2}   } &  9.74 &               1.48 &          133 \\
    \textsl{\textsl{Cluster 3}   } &   9.22 &               1.37 &          148 \\
    \textsl{\textsl{Cluster 4}   } &   8.78 &               5.11 &          162 \\
\bottomrule
\end{tabular}
   \end{center}

    \begin{minipage}{\columnwidth}\small
         Forecaster's average readability score per rationale (comments). CLI: Coleman-Liau Index, ARI: Automated Readability Index.
    \end{minipage}
\label{tab:readability}
\end{table}

\subsection{Comparison with the baseline model}

Lastly, we provide a comparison with a baseline model to assess our clickstream clustering results. To construct the baseline, we cluster crowdworkers by leveraging k-means, using the same 5-grams as feature vectors, such that the comparison with our result is fair and both models are fed the same input data. We select the number of clusters by elbow method, yielding 4 clusters (\textsl{Cluster A}: 236, \textsl{Cluster B}: 63, \textsl{Cluster C}: 47, \textsl{Cluster D}: 212).  Using k-means clustering results as a baseline, we study the same scenarios as before.

\begin{figure}[t]
\centering
  \includegraphics[width=\columnwidth]{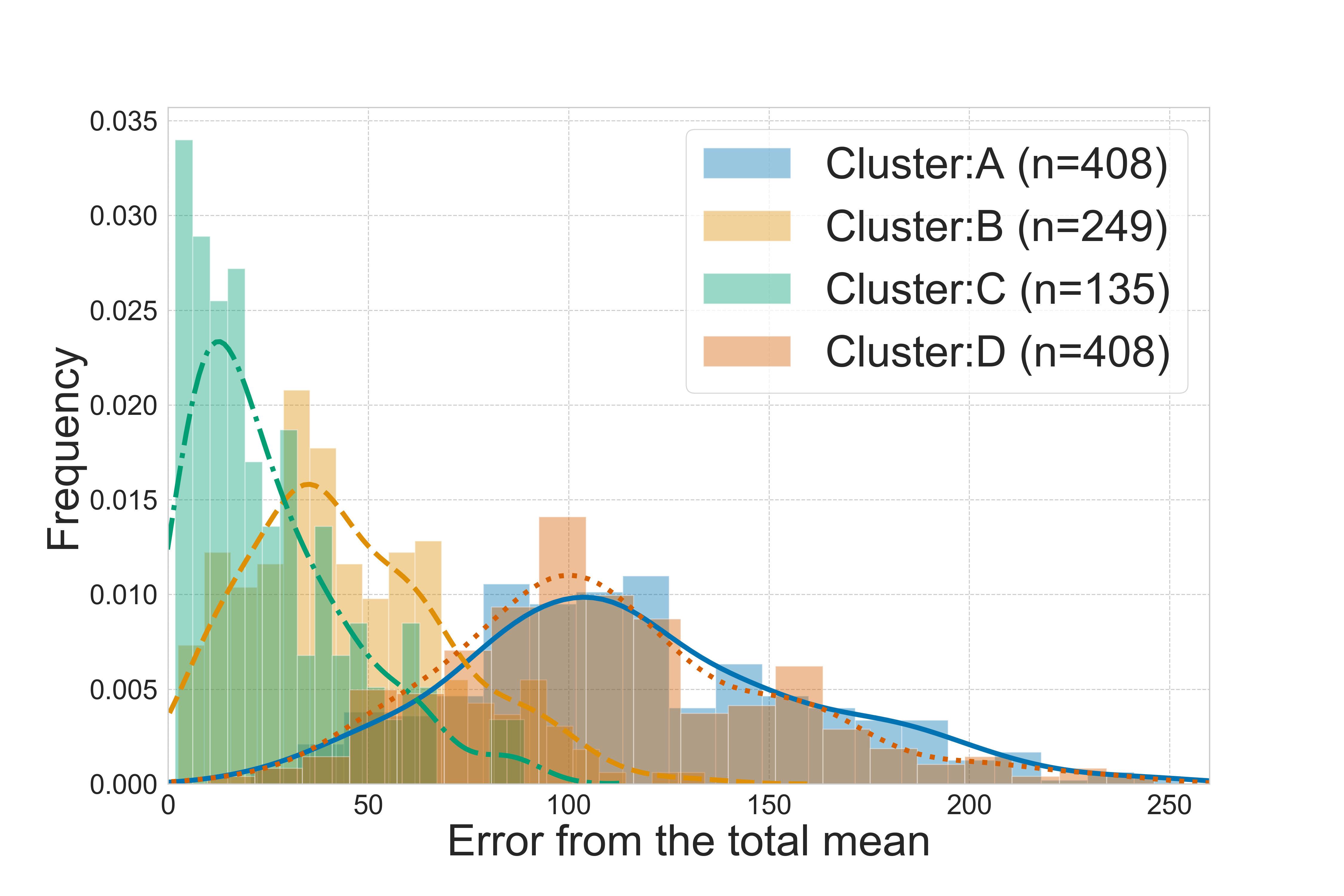}
  \caption{Baseline: Distribution of the Root Mean Square Error (RMSE) between each user's answer and the mean of the answers of other forecasters on a given question.} 
  \label{fig:baseline_error_from_mean}
\end{figure}

\textbf{Scenario 1}: Figure~\ref{fig:baseline_error_from_mean} shows the error from the mean of the baseline model, equivalently to Figure~\ref{fig:error_from_mean}.  The figure shows three groups of distributions. While \textsl{Cluster A} and \textsl{D} show similar distributions, \textsl{Cluster C} and \textsl{D} show skewed distribution with smaller means ($p < 0.001$).  Figure~\ref{fig:baseline_error_from_mean}  shows a spectrum of distributions, therefore it is impossible to select a single candidate low-quality workers cluster as we did in the main analysis.

\begin{figure}[t]
\centering
  \includegraphics[width=0.75\linewidth]{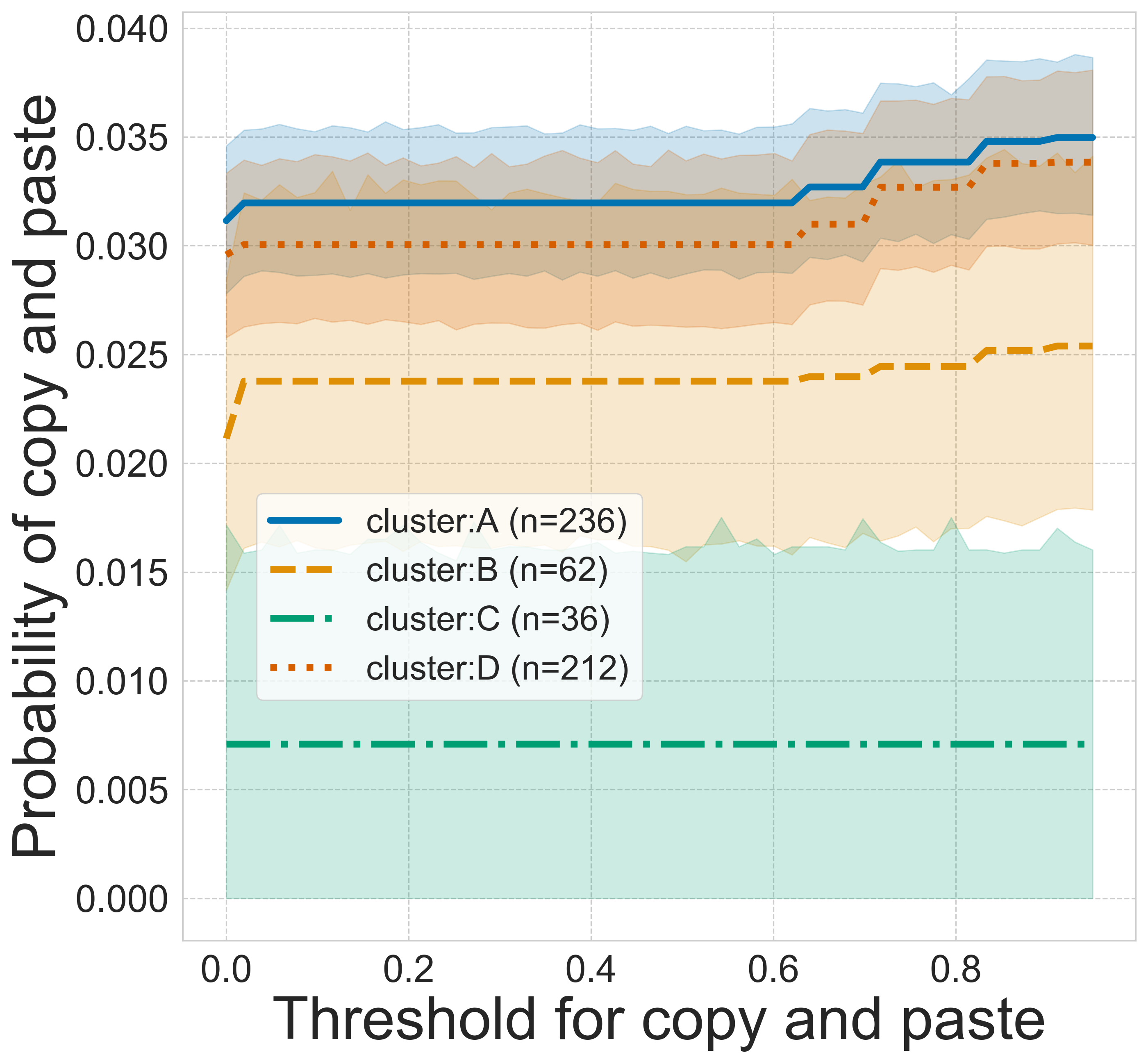}
\caption{Baseline: the transition of the copy and paste per user across the range of the threshold for copy-paste behavior (with 95\% confident intervals).}
    \label{fig:baseline_thr_copy_paste}
\end{figure}

\textbf{Scenario 2}: We could not reject any statistical test for the differences between forecasts and consensus charts (\textsl{Cluster A}: 15.99, \textsl{Cluster B}: 17.90, \textsl{Cluster C}: 19.14, \textsl{Cluster D}: 15.91).  Figure~\ref{fig:baseline_thr_copy_paste} shows the ratio copy-and-paste, equivalently to Figure~\ref{fig:thr_copy_paste}. Figure~\ref{fig:baseline_thr_copy_paste} shows that the error bars of \textsl{Clusters A, C} and \textsl{D} overlap with each other, and \textsl{Cluster C} shows the lowest copy-and-paste ratio. Besides, all clusters show lower copy-and-ratio than \textsl{Cluster 1} in Figure~\ref{fig:baseline_thr_copy_paste}. In summary, we do not see any evidence that a specific cluster shows distinct copy-and-paste behavior in the baseline. 

\textbf{Scenario 3}: We also calculated the four metrics for the baseline clusters. Then, we found that \textsl{Cluster C} writes low-quality rationales. Compared to \textsl{Cluster D}, \textsl{Cluster C} has a small mean CLI readability score (\textsl{Cluster C}: 7.26, \textsl{Cluster D}: 9.28, $p < 0.01$) and shorter text length (\textsl{Cluster C}: 53.94, \textsl{Cluster D}: 76.72, $p < 0.01$). Also, \textsl{Cluster C} write shorter rationales than \textsl{Cluster A} (\textsl{Cluster A}: 72,72, $p < 0.01$). \textsl{Cluster A} also shows fewer misspells per word than \textsl{Cluster D} (\textsl{Cluster A}: 0.111, \textsl{Cluster D}: 0.115, $p<0.05$). The statistical tests for other comparisons are not rejected.

\textbf{Summary.} The analysis of the baseline model yields inconsistent results. Scenario 1 fails to line up a candidate for a low-quality worker cluster. Although Scenario 2 suggests that \textsl{Cluster C} exhibits less copy-paste behavior than other clusters, \textsl{Cluster C} is associated with low-quality rationales in Scenario 3. Altogether, this analysis suggests that our model is significantly more consistent and interpretable than the baseline.

\section{Testing alternative explanations}

We assumed that the clickstream clustering algorithm captures holistic dynamic behaviors that reflect the quality of the forecasters. There may be, however, alternative explanations to the clustering results: for example, the clustering algorithm could depend on some latent factors rather than the sequence of the actions. The first possible alternative could be the case that forecasting prowess tampers clustering results, rather than actual behavioral differences as conveyed by the clickstream trajectories. The other alternative may be that the algorithm picks up the very specific actions related to quality measures such as viewing the consensus charts. In this section, we will test these two  alternative hypotheses.

\subsection{Forecasting prowess}
We verified that the three clusters do not exhibit a statistically significant difference in the mean Brier score of their users: this rejects the first alternative explanation. All pairwise comparisons of the means are statistically indistinguishable ($p > 0.1$) except for \textsl{Cluster 2} VS \textsl{Cluster 4}. The means' difference between \textsl{Cluster 2} VS \textsl{Cluster 4} is statistically significant ($p < 0.01$) but the effect size is less than 1\% difference, a random fluke given the sample sizes. Hence, we reject the first alternative and conclude that the behavioral differences are not caused by differences in users' forecasting prowess, but rather by true differences in clickstream trajectories and associated forecasting behaviors. 

\subsection{Specific behaviors}
Next, we suggest that our clustering does not depend on the variables used in the three scenarios to assess the quality of workers. In Scenario 1, we study the tendency for dispersion of forecasts, which is calculated based on the submitted forecasts. We do not use the forecast values in the clustering. Therefore, we do not expect that the clickstream clustering distorts the assessment of Scenario 1. This is the same for Scenario 3, which investigates the rationales by each forecaster. We do not use any content features of the rationales for clustering. However, in Scenario 2,  we study the copying behavior from consensus charts in which we investigate behavior related to ``consensus-chart'' action. Scenario 2, therefore, needs a more carefully sanity check, discussed next. 

\subsubsection{Is checking the ``consensus-chart'' correlated to copy-and-paste behavior?}

 Our clustering method uses ``consensus-chart'' actions as a part of 5-gram, which might distort the clustering results and our analysis. In an extreme case, for example, if there is a subgroup of the workers who do not use any consensus charts in their clickstream history, we may find a cluster of the users who did not get involved in any ``copying behavior.'' Even though this extreme instance does not happen, the clustering algorithm can pick up users who do not consult the ``consensus-chart'' frequently.
 
 To check if ``consensus-chart'' actions alone are essential variables in the clustering, we study the probability that the workers check a consensus-chart before making forecasts. For a given question, we postulate that a forecast is linked to consulting a consensus-chart if a user viewed the consensus-chart within three actions before making a forecast, compatibly with using 5-grams for clustering.\footnote{Our 5-gram contains time intervals between actions. Therefore, the longest distance between consensus-chart and making forecast action in 5-gram is 3.}
 
 Table~\ref{tab:balance_check_cons} shows \textsl{Cluster 1} and \textsl{Cluster 2} have statistically indistinguishable probability of checking consensus-chart before forecasts ($p > 0.18$). While we conclude that \textsl{Cluster 1} provides more copy-paste forecasts from the consensus chart than \textsl{Cluster 2}, their probability of checking consensus before forecasting is the same. On the other hand, \textsl{Cluster 4} has a lower probability than any other three clusters ($p < 0.01$), but the copy-past behavior of \textsl{Cluster 4} are not different than  \textsl{Cluster 2} and  \textsl{3}. This fact suggests that the probability of checking the consensus charts is not directly related to copy-paste behavior used to assess the quality of the crowdworkers. Finally, we conclude that clickstream clustering does not distort the analysis for Scenario 2.

\begin{table}[t] \small
\caption{Mean Brier score}
   \begin{center}
\begin{tabular}{@{}l@{}@{}ccc@{}@{}}
\toprule
{} & Brier score &  Standard deviation  &  \#Forecasters\\
\midrule
\textsl{Cluster 1} &   0.379  &             0.011 &          104 \\
\textsl{Cluster 2} &  0.381 &               0.011 &          133  \\
\textsl{Cluster 3} &  0.378 &               0.011 &          148 \\
\textsl{Cluster 4} &  0.377 &               0.011 &          162  \\
\bottomrule
\end{tabular}
   \end{center}
    \begin{minipage}{\columnwidth}\small
    The mean value of Brier score for each clusters. The Brier score is the aggregated score based on the answer of each forecasters.
    \end{minipage}
\label{tab:balance_check_brier}
\end{table}

\begin{table}[t] \small
\caption{Consensus chart check }
\begin{tabular}{@{}l@{}@{}ccc@{}@{}}
\toprule
{} & Consensus chart check & Standard deviation  &  \#Forecasters\\
\midrule
\textsl{Cluster 1} &  0.965 &               0.042 &          104 \\
\textsl{Cluster 2} &  0.966 &               0.044 &          133  \\
\textsl{Cluster 3} &  0.959 &               0.046 &          148 \\
\textsl{Cluster 4} &  0.864 &               0.220 &          162  \\
\bottomrule
\end{tabular}

    \begin{minipage}{\columnwidth}\small
    The probability of checking the Consensus chart before making forecasts for each cluster. The probability is calculated as the average ratio that the users view the Consensus chart within 3 actions before making forecasts.
    \end{minipage}
\label{tab:balance_check_cons}
\end{table}

\section{Related Work}

\subsection{Quality of Crowdworkers}

Crowdworkers are often used to construct the data set for research such as annnotation~\cite{finin2010annotating,nowak2010new}, question, and answer dataset~\cite{sheehan2018crowdsourcing,buhrmester2011amazon,mason2012conducting}. Low-quality workers may threaten the results and validity of the research. Hence, studying the validity of crowdworkers has been a crucial issue. Assessments of crowdworker quality can be straightforward when the clowdworkers work in the tasks with the groundtruth such as annotation tasks~\cite{nowak2010reliable,gillick2010non,yetisgen2010preliminary}. However, in many cases, constructing groundtruth data costs a lot. In our context of forecasting tasks, it is impossible to have a groundtruth label beforehand. To overcome this problem, many studies have assessed the quality of crowdworks without groundtruth.

In the literature, the assessment of the quality of workers takes three forms. The most popular practice is comparing the results of a study with workers to the existing research. This form of assessments, for example, has been conducted in studies with human subject like laboratory experiments,~\cite{paolacci2010running,horton2011online,suri2011cooperation,johnson2012participants,goodman2013data}, or surveying studies \cite{rouse2015reliability,goodman2013data,holden2013assessing}. Also, a meta-analysis of the existing studies with crowdworkers can provide high-level assessment for the consistency of crowdworkers~\cite{mortensen2018comparing}. Conducting peer-reviewing can be in this form of assessment~\cite{tang2019leveraging}.

In the second group, the external information about the crowdworkers is used as a clue to learn the quality of crowdworkers. The consistency between the estimated worker's location and their self-report living place can validate the workers~\cite{buhrmester2011amazon}. Reputations or consistency Comparing the results by the low reputation crowdworkers with the high reputation ones~\cite{peer2014reputation}.

The last group of the literature utilizes the behavior trajectory of the crowdworker to assess the quality of their work. Since crowdworkers supply their labor in virtual platforms, every single move of individual workers can be traced at a small cost. The trace of crowd worker behavior, for example, can be used to predict the workers' quality by a supervised machine learning model~\cite{rzeszotarski2011instrumenting,han2016crowdsourcing}. Their findings imply that behavior traces tell the quality of the workers. In a similar fashion, clustering the worker behavior can group the workers that share a similar quality of works, for example, in annotations tasks~\cite{kairam2016parting}.

While the last group of the literature provides a way to understand the quality of corwdworkers, they have two limitations. First, their assessments are conducted on the crowd tasks with groundtruth. Some essential tasks solved in crowdsourcing do not have "correct" labels, for example, forecasting, surveying, prediction markets, etc. Also, their machine learning models do not incorporate the dynamics of behaviors. For instance, while the amount of time spend on tasks are used as features~\cite{rzeszotarski2011instrumenting,han2016crowdsourcing}, sequences of the behavior are not used. 

Our paper aims at solving the issues in the last group in the literature outlined above. We study the forecasting tasks, which have no correct answer when they are assigned. Instead of using the accuracy of the forecasts, we use behaviors that do not directly relate to the tasks. We then detail the submitted answers by forecasters to assess their quality. In addition, to study the temporal feature of workers' behavior, we use the clickstream clustering, where sequences of the workers' behavior are used as user features.

\subsection{Clickstreaming analysis}

Our work builds upon a wealth of previous research on clickstream clustering~\cite{srivastava2000web,lu2005mining,sadagopan2008characterizing,benevenuto2009characterizing}. In our analysis, we use the unsupervised clickstream clustering framework from Wang and collaborators~\cite{wang2016unsupervised}.  
This and other similar frameworks are postulated upon the intuition that the sequence of actions that a user performs to accomplish a task is indicative, and often predictive, of typical patterns of behaviors. In turn, these can be used to separate users into groups exhibiting similar characteristics. Behavioral clustering based on clickstream data has seen a wealth of applications in Web and online user behavioral modeling~\cite{gunduz2003web,ting2005ubb,wang2013you,su2015method}.  Wang et al. \cite{wang2013you}, for example, distinguish between users with different goals (in their case, social media sybil and genuine accounts). Our approach, in contrast, aims to identify low-quality workers among the users with the same goal (completing the forecast tasks). In this paper, we provide empirical evidence that clickstream clustering can be used to detect low-quality workers in crowdsourcing platforms.

\section{Conclusions}
In this work, we study the broad problem of understanding user behavior based upon their clickstream. Specifically, we identify low-quality workers by clustering clickstreams of forecasts on a geopolitical forecasting platform. Using a state-of-the-art clickstream clustering approach, we find that we are able to identify groups of users who simply adopt their forecasts by viewing the aggregate consensus of their peers, adding no meaningful information into the system whatsoever. We found that one cluster has a preponderance of copying \emph{exactly} the same forecast as is found in the consensus chart shown to all workers of our platform. Through additional inspection, we find that specific clickstream behaviors, such as only viewing ratings and the consensus chart, are indicative of this behavior. 

While our study focuses on forecasting behavior, it has implications for the wider field of crowdsourcing. Our methodology demonstrates how crowdsourcing tasks can estimate low-quality workers and results by observing the redundancies in the shared behavior of these users. 

Our method can reduce redundancy (thus saving money), improve user behavior, and prediction accuracy. Low-quality crowdworkers can be invited less frequently to future tasks - however, they are always paid the same as others. Such workers can also be targeted for behavioral interventions, like additional training to help them improve and become better forecasters. We also want to emphasize the paramount importance not to share the behavior traces without permission with other platformers to protect the privacy of the crowd workers.

Future work is to extend the methodology to conform to the nuances of geopolitical forecasting, and to develop computational frameworks that can reason over a lack of observed behavior during a task. For instance, when a user is not generating actions on the crowdsourcing platform, it is not clear whether they are simply idle or they are doing research in Web another tab (e.g., reading related news articles). Future work is building statistical approaches to better assess this behavior.

\small \textbf{Acknowledgements}. This work is supported by DARPA (grant \#D16AP00115) and IARPA (via 2017-17071900005).

\end{document}